\documentclass[aps,prl,twocolumn,showpacs,groupedaddress]{revtex4}
\usepackage{graphicx}
\usepackage{amssymb}
\usepackage{amsmath}
\usepackage{xcolor}
\newlength{\figwidth}

\usepackage{epsf}
\begin{document}
\preprint{\today}
\title{Scanning Gate Microscopy of Kondo Dots: \\
Fabry-P\'erot Interferences and Thermally Induced Rings}

\author{Andrii Kleshchonok}
\affiliation{Service de Physique de l'\'Etat Condens\'e (CNRS URA 2464), 
IRAMIS/SPEC, CEA Saclay, 91191 Gif-sur-Yvette, France}

\author{Genevi\`eve Fleury}
\affiliation{Service de Physique de l'\'Etat Condens\'e (CNRS URA 2464), 
 IRAMIS/SPEC, CEA Saclay, 91191 Gif-sur-Yvette, France}

\author{Jean-Louis Pichard}
\affiliation{Service de Physique de l'\'Etat Condens\'e (CNRS URA 2464), 
IRAMIS/SPEC, CEA Saclay, 91191 Gif-sur-Yvette, France}

\begin{abstract} 
We study the conductance of an electron interferometer formed in a two dimensional electron gas  
between a nanostructured quantum contact and the charged tip of a scanning gate microscope. Measuring 
the conductance as a function of the tip position, thermally induced rings may be observed in addition 
to Fabry-P\'erot interference fringes spaced by half the Fermi wavelength. If the contact is made of a 
quantum dot opened in the middle of a Kondo valley, we show how the location of the rings allows to 
measure by electron interferometry the magnetic moment of the dot above the Kondo temperature. 

\end{abstract}  

\maketitle

\begin{figure}
\includegraphics[keepaspectratio,width=\columnwidth]{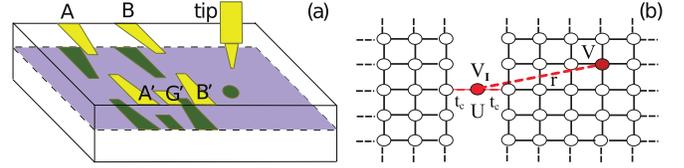}
\caption{(Color online) (a): Metallic gates (yellow) create depletion regions (green) in the 2DEG (blue) beneath the surface. This 
makes a dot with a gate in the contact between left and right 2DEGs, while a charged tip yields a scannable depletion region near the contact. 
The SGM images give the conductance as a function of the tip position. (b): Two semi-infinite square lattices are contacted via an 
Anderson impurity (site ${\bf I}$ of coordinates $(0,0)$, Hubbard repulsion $U$, potential $V_{\bf I}$, coupling $t_c$). Adding a 
potential $V$ at a site of coordinates $(x,y)$ gives rise to an electron interferometer of size $r$.  
} 
\label{fig1}
\end{figure}

 Scanning gate microscopy (SGM) is a new tool which allows to probe by electron interferometry~\cite{Topinka:PT03} the properties of nanostructures 
created in a two-dimensional electron gas (2DEG). The nanostructures are made with charged gates deposited 
on the surface of a semi-conductor heterostructure, allowing to divide the 2DEG beneath the surface in two parts connected via a more or 
less simple contact region. This can be a quantum point contact~\cite{PhysRevLett.60.848,PhysRevLett.77.135} (QPC), a quantum 
dot~\cite{nature391,science281,Goldhaber-gordon}, a double dot setup~\cite{Molenkamp:PRL95} or more complex nanostructures. With five 
gates, the contact between the right and left leads (left and right parts of the 2DEG) can be made of a quantum dot with a tunable gate 
(see Fig.~\ref{fig1}(a)). With the charged tip of an atomic force microscope above the surface of the heterostructure, a depletion region can be 
capacitively induced in the 2DEG below the surface at a distance $r$ from the contact. SGM consists in studying the conductance $g$ of the 
electron interferometer formed by the contact and the depletion region. Scanning the tip outside the contact, one can record SGM images giving 
$g$ as a function of the tip position. These images exhibit Fabry-P\'erot interference fringes spaced by $\lambda_F/2$, as observed by 
Topinka et al~\cite{Topinka:Sci00} for a QPC opened on its first conductance plateau. Using high-mobility 2DEGs, the SGM images of a single 
QPC have been more systematically investigated later at low temperatures in Refs.~\cite{Topinka:Nat01,LeRoy:PRL05,Jura:PRB09,ensslin} for 
different points of a conductance plateau as well as between plateaus. This has led to revisit the theory of electron 
interferometers~\cite{Heller:NL05,metalidis,fkp,Jalabert:PRL10,alp,gorini} which include a quantum point contact.  Recently, the spacing 
between the interference fringes at a certain distance from the contact was found~\cite{ensslin} to differ by more than $50 \% $ from the expected 
value $\lambda_F/2$ when the QPC is biased on the second, third and fourth conductance plateaus. This difference gives rise to a large ring of 
radius $\approx 1 \mu m$ visible in the SGM images of Ref.~\cite{ensslin} where different scenarii were proposed for explaining this unexpected 
ring. This leads us to study if interference mechanisms other than those responsible for the $\lambda_F/2$-oscillations can occur in the limit 
where the electron motion is purely ballistic between the contact and the tip. We show in this letter that this can indeed occur for a contact 
characterized by a series of transmission peaks, if it is opened between two peaks. Moreover, for a dot with an odd number of electrons and 
biased in the middle of the Kondo valley, this opens the possibility to measure by electron interferometry the magnetic 
moment~\cite{nature391,science281,Goldhaber-gordon} induced by electron-electron interactions above the Kondo temperature. 

 In the Coulomb blockade regime, the gate voltage can be tuned for having an odd number of electrons in a lithographically defined quantum dot. 
This gives rise to an unpaired spin. Our study is based on the model routinely used~\cite{Goldhaber-gordon} for describing the Kondo effect due 
to this unpaired spin: An Anderson impurity coupled to two semi-infinite square lattices, the many body effects coming from the presence 
of an Hubbard repulsion $U$ in the contact (see Fig.~\ref{fig1}(b)). The Hamiltonian without tip reads $H_{0}=H_{\bf I}+H_c+H_{leads}$ 
where  
\begin{eqnarray}
&H_{\bf I}= V_{\bf I} \sum_{\sigma} n_{{\bf I}\sigma} + U n_{{\bf I}\uparrow}n_{{\bf I}\downarrow} \\
&H_c= t_c \sum_{\sigma} \left(c^{\dagger}_{{\bf I}\sigma} c_{{(0,1)}\sigma}+c^{\dagger}_{{\bf I}\sigma} c_{{(0,-1)}\sigma}+ H.C.\right) \\  
&H_{leads}= \sum_{{\bf i}\sigma} \left( -4t n_{{\bf i}\sigma} + t \sum_{\bf j} c^{\dagger}_{{\bf i}\sigma} c_{{\bf j}\sigma} \right) + H.C. 
\label{}
\end{eqnarray}
$c_{{\bf i}\sigma}$ ($c^{\dagger}_{{\bf i}\sigma}$) is the destruction (creation) operator of an electron of spin $\sigma$ at site ${\bf i}$ and 
$n_{{\bf i}\sigma}=c^{\dagger}_{{\bf i}\sigma}c_{{\bf i}\sigma}$. $H_{\bf I}$ describes the Anderson impurity located at the site ${\bf I}$ of 
coordinates $(0,0)$ making the contact and $H_c$ the coupling to the leads (hopping $t_c$ between ${\bf I}$ and the two neighboring sites of the 
two leads). $H_{leads}$ describes two semi-infinite 
square lattices making the right and left leads (nearest neighbor hopping $t$). 
The energy scale is defined by taking $t=-1$ and site potentials equal to $-4t$ yield energy bands $[0,8]$ for the conduction electrons of the leads. 
Hereafter, we study the continuum limit and consider only small energies $E$. To this model for the contact, we add a term $H_{tip}=\sum_{\sigma} V 
n_{{\bf T}\sigma}$, assuming that the depletion region induced by the charged tip modifies only the potential of a single site ${\bf T}$ of coordinates 
$(x,y)$ located at a distance $r$ from the contact. The interferometer Hamiltonian reads $H=H_{0}+H_{tip}$. 

{\it Dot without interaction:}  When $U=0$, this model can be analytically solved~\cite{alp,lp}. Let us summarize the main results (partly given in 
Ref.~\cite{alp} and with more details in Ref.~\cite{lp}) which are necessary for understanding how a pattern of interference rings with a period 
different from $\lambda_F/2$ can be seen in the SGM images. The contact being reduced to a single site ${\bf I}$ coupled to another single site per 
lead, the lead self-energies~\cite{Datta:book97} are only two complex numbers 
$\Sigma_{l,r}=R_{l,r}(E)+iI_{l,r}(E)=t_c^2<\pm 1,0|G_{l,r}^R(E)|\pm 1,0>$, $G^R_{l,r}(E)$ being the retarded Green's function of the left 
and right leads evaluated at the sites directly coupled to ${\bf I}$. Using the method of mirror images~\cite{Molina:PRB06}, 
$G^R_{l,r}(E)$ can be expressed in terms of the Green's function $G^R_{2d}(E)$ of the infinite 2d lattice~\cite{Economou:book06}. 

Without tip ($V=0$), the transmission $T^{\sigma}_{0}(E)$ of an electron of spin $\sigma$ through the dot making the contact reads: 
\begin{equation}
T_0^{\sigma}(E)= \frac{4 I_rI_l}{(E-4-V_{\bf I}-R_r-R_l)^2+(I_r+I_l)^2}.     
\label{transmission_without_tip}
\end{equation}
If the variation of $\Sigma_{l,r}(E)$ can be neglected when $E$ varies inside the resonance (typically $t_c<0.5$ in the continuum limit 
where the Fermi momentum $k_F \ll 1$), this is a Lorentzian of width $\Gamma = -(I_r+I_l)=-2I$ and center $4+V_{\bf I}+2R$ if $R_l=R_r=R$. 

If one adds a tip potential $V \neq 0$ in the right lead, the effect of the tip can be included by adding an amount 
$\Delta \Sigma_r(E)=\Delta R_{r}(E)+i\Delta I_{r}(E)$ to $\Sigma_r(E)$ (see Refs.~\cite{alp,lp,Darancet:PRB10}). The 
interferometer transmission $T^{\sigma}(E)$ is still given by Eq.~\eqref{transmission_without_tip}, once $R_r+\Delta R_r$ and 
$I_r+\Delta I_r$ have been substituted for $R_r$ and  $I_r$. Moreover, when $r$ is sufficiently large, $\Delta \Sigma_r(E,r)$ becomes small 
and one can expand $T^{\sigma}(E)$ in powers of $\Delta \Sigma_r(E)$. The effect of the tip being restricted to a single site ${\bf T}$, 
$\Delta \Sigma_r$ can be obtained from Dyson's equation. In the continuum limit and for distances $r \gg k_F^{-1}$, one finds~\cite{alp,lp} 
\begin{equation}   
\frac{\Delta \Sigma_r}{t_c^2 \rho} \approx -\frac{k_F x^2}{2 \pi r^3} \exp[i(2k_Fr+\pi/2+\phi)] + O(\frac{x^{3/2}}{r^3}),   
\label{self-energy}
\end{equation} 
where $\rho$ and $\phi$ are the modulus and the phase of $V/(1-V \langle 0,0|G^R_{2d}(E)|0,0\rangle )$. Expanding $T^{\sigma}(E)$ to the leading order 
$\propto x^2/r^3$ in $\Delta \Sigma_r$, the effect of the tip upon the conductance at a temperature ${\cal T}$ can be obtained:
\begin{equation}
\Delta g=g-g_0= \sum_{\sigma}\int dE  (T^{\sigma}(E)-T^{\sigma}_{0}(E)) (-\frac{\partial f}{\partial E}),
\label{conductance-change}
\end{equation} 
where $f$ is the Fermi-Dirac distribution. Assuming~\cite{Heller:NL05} $-\partial f/ \partial E \approx (1/4k_B{\cal T}) 
\exp-[\sqrt{\pi}(E-E_F)/(4k_B{\cal T})]^2$ and approximating $\Sigma_{l,r}$, $\rho$, and $\phi$ by their values at 
the Fermi energy $E_F$, one eventually gets~\cite{lp}:  
\begin{eqnarray}
&\Delta g ({\cal T})\approx 2 A({\cal T}) \cos (2k_Fr+\Phi({\cal T})), \\
&A({\cal T})=\frac{\rho x^2 l_{{\cal T}}}{\sqrt{\pi}k_F r^3} \exp -[(1+v^2)(\frac{l_{\cal T}}{l_{\Gamma}})^2 +\frac{r}{l_{\Gamma}}] \\
&\Phi({\cal T})=\phi+v\frac{r}{l_{\Gamma}}-2v(\frac{l_{\cal T}}{l_{\Gamma}})^2. 
\label{conductance}
\end{eqnarray}
$v \equiv (V^{res}_{\bf I}-V_{\bf I})/\Gamma$ gives the energy shift of $V_{\bf I}$ from the resonance $V^{res}_{\bf I}\equiv E_F-4-2R$ in units of 
$\Gamma$.  $l_{\cal T}=(\sqrt{\pi}k_F)/(4k_B{\cal T})$ and $l_{\Gamma}=k_F/\Gamma$ are two length scales associated respectively to ${\cal T}$ 
(Fermi-Dirac statistics) and to $\Gamma$ (resonant transmission). To obtain Eq.~\eqref{conductance}, we have used asymptotic expansions valid when 
$r>r^*\equiv 2 l_{\cal T}[1+l_{\cal T}(1+|v|)/l_{\Gamma}]$. The factor $2$ in  $\Delta g ({\cal T})$ comes from the spin degeneracy.

\begin{figure}
\includegraphics[keepaspectratio,width=\columnwidth]{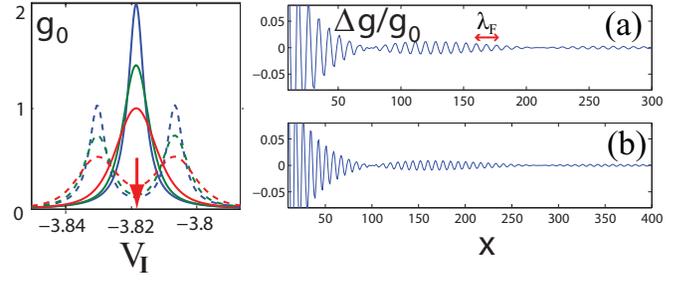}
\caption{(Color online) Left: Conductance $g_0$ of a non interacting dot without tip ($V=0$, $t_c=0.2$, and $\Gamma=0.003$) as a function of 
$V_{\bf I}$ for ${\cal T}=0$ (blue) ${\cal T}=\Gamma/2$ (green) and ${\cal T}=\Gamma$ (red). Cases with ($h=0$, solid line) and without spin 
degeneracy ($h=4\Gamma$, dashed line). The arrow gives the potential $V_{\bf I}$ used in the right figures. Right: $\Delta g/g_0 
({\cal T}^*,\Gamma^*,y=0)$ as a function of $x$ for a tip potential $V=-2$. (a): $h=0.0091$, ${\cal T}^*=0.0066$, $\Gamma^*=0.0023$; (b): 
$h=0.0068$, ${\cal T}^*=0.005$, $\Gamma^*=0.0017$. In all figures, $E_F=0.1542$ and $\lambda_F/2=8$.}
\label{fig2}
\end{figure}

{\it Zeeman splitting without interaction:}  Let us consider now the case where the spin degeneracy is removed by a Zeeman term $\pm h$ 
due to a parallel magnetic field applied in the dot. This removal is illustrated in Fig.~\ref{fig2} (left), $g_{0}^{\uparrow}({\cal T})$ having 
a peak shifted by an amount $v^{\uparrow}=-h/\Gamma$ (electron with parallel spin) while the shift is $v^{\downarrow}=h/\Gamma$ for the antiparallel 
spin. Hereafter, we study the effect of the tip at the value $V^{res}_{\bf I}(h=0)$ for $V_{\bf I}$ indicated by an arrow in Fig~\ref{fig2} left 
(middle of the valley). When $h \neq 0$, $\Delta g ({\cal T})=\sum_{\sigma} A({\cal T}) \cos (2k_Fr+\Phi^{\sigma}({\cal T}))$. The amplitudes $A$ 
depending on $v^2$ are independent of $\sigma$, but the phases $\Phi^{\sigma}$ depend on the sign of $v$, and hence of $\sigma$. 
%This is the usual phase shift of $\pi$ of the transmission amplitude when the energy crosses a resonance. 
This gives rise to a beating effect, the oscillations of  $\Delta g^{\uparrow}({\cal T})$ and $\Delta g^{\downarrow}({\cal T})$ canceling each other 
and the tip having no effect on $g({\cal T})$ when the distance $r$ takes values 
\begin{equation} 
r^D(n)=\frac{2k_F}{\Gamma}(\frac{l_{\cal T}}{l_{\Gamma}})^2+(n \pi +\frac{\pi}{2}) \frac{k_F}{h}, 
\label{radius-ring}
\end{equation}   
where $n$ is a positive integer. Conversely, the oscillations of  $\Delta g^{\uparrow}({\cal T})$ and $\Delta g^{\downarrow}({\cal T})$ add 
if $r(n)=r^D(n)+\pi k_F/(2h)$. The SGM image giving $\Delta g ({\cal T})$ as a function of the tip position is characterized by a first ring at a distance 
$r^D(n=0)$ followed by other rings spaced 
by $\pi k_F/h$ where $\Delta g ({\cal T})=0$. To optimize the contrast in the images, we calculate for a given value of $h$ the temperature ${\cal T}^*$ 
and the width $\Gamma^*$ for which $\Delta g({\cal T},r=r^D(n=0)+\pi k_F/(2h))$ is maximum. The extrema are given by two coupled non-linear algebraic 
equations which can be solved numerically, yielding ${\cal T}^* \approx 0.73 h$ and $\Gamma^*\approx 0.25 h$. In Figs.~\ref{fig2} (a) and (b), 
$\Delta g({\cal T}^*,\Gamma^*)$ is shown as one varies the tip at the right side of the contact, keeping the tip coordinate $y=0$. The figures 
correspond to two values of $h$ chosen using Eq.~\eqref{radius-ring} and the conditions $r^D(h,n=0)=75$ and $100$. The numerical results give rings at 
the expected distances $r^D(n=0)$ and $r^D(n=1)$. Though the period of the oscillations is $\lambda_F/2$, the oscillations around  $r \approx r^D(n)$ 
become so small that one can easily miss a few of them, and draw the conclusion that the period of the oscillations exceeds $\lambda_F/2$.
    
{\it Anderson impurity making the contact:} 
 Let us now consider the case where the resonance peak of $g_0(E)$ is not split in two peaks by an applied magnetic field, but by an Hubbard 
repulsion of strength $U$ acting in the contact. This case is of particular interest, since it describes a quantum dot with an odd number of 
electrons. This splitting was observed by measuring~\cite{nature391,science281,Goldhaber-gordon} the dot conductance as a function of an 
applied gate voltage $V_g$. The interval between the two conductance peaks is called the Kondo valley. It vanishes below the Kondo temperature, 
the dot becoming transparent (unitary limit) in the interval of $V_g$ where the number of electrons remains odd. Hereafter, we study the interferometer 
conductance when the dot is biased in the middle of the Kondo valley, describing the resonant level of the dot by the Anderson model. The middle of the 
Kondo valley corresponds to the symmetric case where $V_{\bf I}+(4+2R)=E_F-U/2$. When $U > \Gamma$, it exhibits three fixed points as the temperature 
${\cal T}$ decreases~\cite{krishna-murthy2,tsvelick-wiegmann1}. At large temperatures (${\cal T} > \sqrt{2 U \Gamma}/\pi$), the Anderson impurity coupled 
to the conduction electrons (left and right 2DEGs) is described by the excitations of the free orbital fixed point. For an intermediate range of 
temperature ($\sqrt{2U \Gamma}/\pi>{\cal T}>{\cal T}_K$), the impurity has a local magnetic moment and the system excitations become different (local moment 
fixed point). Below the Kondo temperature ${\cal T}_K=(\sqrt{2U\Gamma}/\pi) \exp-[\pi U/(8 \Gamma)]$, the local moment is screened by the conduction electrons 
and the excitations are those of the strong coupling limit. If the occurrence of a magnetic moment can be detected using the Hartree-Fock (HF) 
approximation~\cite{anderson}, more involved many-body methods as the numerical renormalization group (NRG) algorithm~\cite{krishna-murthy2} or the Bethe 
ansatz~\cite{tsvelick-wiegmann1} are necessary to describe the Kondo screening of the magnetic moment. Hereafter, we study the effect of the tip for 
temperatures above ${\cal T}_K$ using the HF approximation. Though this can be questionable in the magnetic region, where a HF-description 
can give artifacts, it is usually believed that the HF-behaviors are very suggestive, and, when suitably reinterpreted, indicate what we can expect an exact 
treatment to yield~\cite{Stewart,haldane}. Notably, the HF-approximation does not give the spin-flip processes~\cite{logan}. This point is extremely important. 
If the local moment has a finite time of flip $\tau_{sf}$, the conductance oscillations cannot persist beyond a coherence length $l_{\phi}=k_F\tau_{sf}$, and a 
mean-field theory breaks down on scales $r > l_{\phi}$. Nevertheless, the oscillations should be given by the HF-approximation on shorter scales. 
$\tau_{sf}$ can be estimated for the Anderson model using NRG~\cite{krishna-murthy2} or Bethe ansatz~\cite{tsvelick-wiegmann1}. From the 
expression giving the magnetic susceptibility as a function of the magnetic moment $\mu$, ${\cal T}$ and ${\cal T}/{\cal T}_K$~\cite{krish-murthy1,
filgov-tsvelick-wiegmann,andrei-lowenstein}, one can estimate $\tau_{sf}$ as a function of ${\cal T}$. For the temperature 
range $\sqrt{2U \Gamma}/\pi>{\cal T} \gg {\cal T}_K$ which we shall consider, $l_{\phi} \gg (l_{\cal T},l_{\Gamma})$   
($l_{\phi}\approx 1000$ when ${\cal T}/{\cal T}_K\approx 30$). 

\begin{figure}
\includegraphics[keepaspectratio,width=\columnwidth]{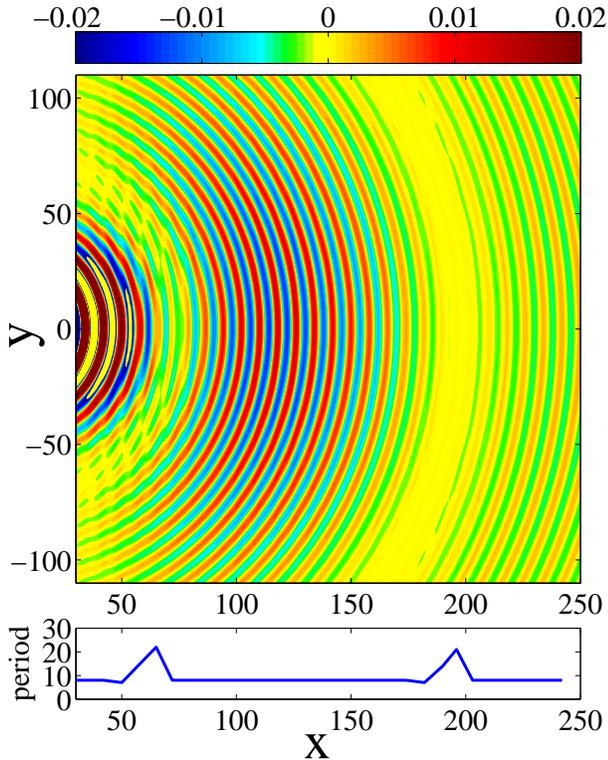}
\caption{(Color online)  $\Delta g/g_0$ (upper color scale) as a function of the tip coordinates $(x,y)$ when an Anderson impurity with $U=0.0316$ 
and $t_c=0.1592$ makes the contact. $E_F=0.1542$, $V=-2$ and $g_0=0.2607$. The temperature (${\cal T}^*\approx 0.0054 \gg {\cal T}_K$) and the resonance 
width ($\Gamma^*\approx 0.0019$) have been calculated for having the largest amplitude $\Delta g/g_0$ after $r^D(n=0)=75$. The apparent period between 
the visible maxima of $\Delta g/g_0$ is given in the lower figure. It is equal to $\lambda_F/2=8$, but becomes larger than $\lambda_F/2$ around 
$r^D(n=0)=75$ and $r^D(n=1)=195$. 
}
\label{fig3}
\end{figure}

\begin{figure}
\includegraphics[keepaspectratio,width=\columnwidth]{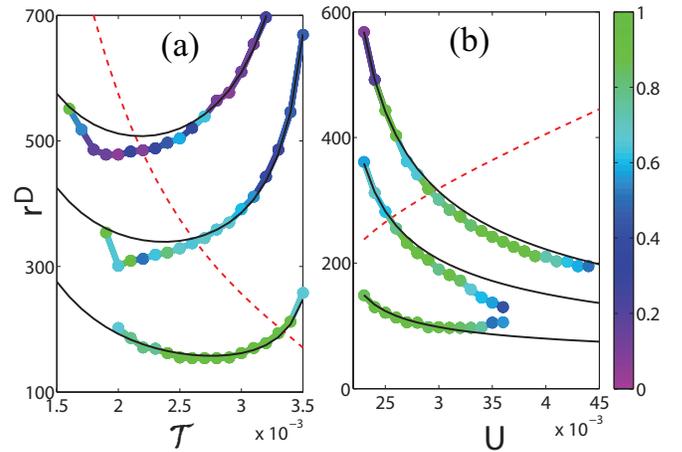}
\caption{(Color online) 
Radii $r^D(n)$ of the rings for $n=0, 1$ and $2$ as a function of the temperature ${\cal T}$ (left, $U=0.022$) and of the interaction strength 
$U$ (right, ${\cal T}=0.0032$) with $\Gamma=0.003$. The curves are in the regimes ${\cal T}_K \ll {\cal T} < \sqrt{2U\Gamma}/\pi$ and $U> \Gamma$. The dots 
are obtained numerically, their colors corresponding to a visibility scale~\cite{born} indicated at the right. The solid lines are the analytical values 
of $r^D(n)$ derived assuming $r^D(n)>r^*$ (dashed lines). One can see the ranges of temperature and interaction where the first ring is sufficiently close 
to the contact for being visible in the SGM images. 
}
\label{fig4}
\end{figure}

In a mean-field approximation, the impurity potential $V_{\bf I}$ is corrected by real Hartree potentials $\Sigma^H_{\sigma}$, 
given by the self-consistent solutions of coupled equations:
\begin{eqnarray}
&\Sigma^H_{\downarrow}=-\frac{U}{\pi} \int dE f(E,E_F,{\cal T}) \Im G^R_{\uparrow}(E) 
\label{HF-eqs-1} \\
&\Sigma^H_{\uparrow}=-\frac{U}{\pi} \int dE f(E,E_F,{\cal T}) \Im G^R_{\downarrow}(E),
\label{HF-eqs-2}
\end{eqnarray}   
where $f$ is the Fermi distribution at temperature ${\cal T}$ and Fermi Energy $E_F$ and 
$G_{\sigma}(E)=[E-V_{\bf I}-4-\Sigma_l-\Sigma_r-\Delta\Sigma_r-\Sigma^H_{\sigma}]^{-1}$ is the HF Green's function of an electron of energy $E$ 
and spin $\sigma$. Without tip, this gives a splitting of the resonance by an amount $U|n_{{\bf I}\uparrow}-n_{{\bf I}\downarrow}|$ proportional 
to the interaction induced magnetic moment. Such a magnetic moment can be modified by the tip. However, the effect of the tip upon $\Sigma^{H}_{\sigma}$ 
remains negligible for the values of $r$ and ${\cal T}$ used in this study. It was shown previously for a similar model that the tip modifies the 
HF potentials by an amount $\Delta \Sigma^{HF}$ which decays as $1/r^2$ at zero temperature~\cite{fkp}. 
In the Anderson model, we obtain~\cite{kfp2} that $\Delta \Sigma^{H}_{\sigma} \propto (1/r^{\alpha})(\exp-[r/l_{\cal T}])$ at a temperature ${\cal T}$. 
Though the exponent $\alpha$ of the decay can slightly differ when the resonance is narrow ($\alpha \to 1$ when $t_c \to 0$), the exponential damping makes 
the correction quickly negligible. Neglecting the effect of $\Delta\Sigma_r$ upon $\Sigma^H_{\sigma}$, the dot biased in the middle of a Kondo 
valley can be described by the theory previously developed for the non interacting dot, with a magnetization which is not now a free parameter 
induced by an external field, but takes a self-consistent value which depends on ${\cal T}$, $U$ and $\Gamma$. The corresponding SGM images should 
also exhibit interference rings, characterized by radii $r^D(n)$ given by Eq.~\eqref{radius-ring} where one has substituted the self-consistent value 
of $U|n_{{\bf I}\uparrow}-n_{{\bf I}\downarrow}|$ for $2h$. The rings are now spaced by a distance $2\pi k_F/(U|n_{{\bf I}\uparrow}-n_{{\bf I}\downarrow}|)$. 
This makes possible to measure the magnetic moment by electron interferometry. To optimize the contrast of the SGM images, we calculate for a given 
value of $U$ the temperature ${\cal T}^*$ and the width $\Gamma^*$ for which 
$\Delta g ({\cal T},r=r^D(n=0)+\pi k_F/(U |n_{{\bf I}\uparrow}-n_{{\bf I}\downarrow}|))$ is maximum. 
The extrema are now given by two coupled self-consistent differential equations, which can be solved numerically, yielding ${\cal T}^* \approx 0.17 U$ 
and $\Gamma^*\approx 0.06 U$. In Fig.~\ref{fig3}, the SGM image of a dot biased in the middle of a Kondo valley is shown for a set of optimized values 
${\cal T}^*$ and $\Gamma^*$. One can see two rings in addition to the Fabry-P\'erot oscillations. Near the rings, the apparent increase of the spacing 
between the interference fringes (see lower part of Fig.~\ref{fig3}) is reminiscent of the effect reported in Ref.~\cite{ensslin} obtained with a QPC 
instead of a Kondo dot. In Fig.~\ref{fig4}, the radii $r^D(n)$ of the rings are given as a function of 
${\cal T}$ (\ref{fig4}(a)) and $U$ (\ref{fig4}(b)) for a set of fixed values of $U$ and $\Gamma$ or ${\cal T}$ and $\Gamma$ respectively. 
The right color scale gives a visibility parameter~\cite{born} equal to $0$ without contrast and to $1$ with a perfect contrast. 
The radii $r^D(n)$ obtained from the numerical self-consistent solutions of Eqs.~\eqref{HF-eqs-1} and \eqref{HF-eqs-2} (without neglecting the effect of 
$\Delta\Sigma_r$ in the HF-potentials) turn out to be well approximated by our simplified theory (which neglects it). When ${\cal T} \to {\cal T}_K$, 
the radii $r^D(n)$ become too large for observing rings. When  ${\cal T} \to \sqrt{2U\Gamma}/\pi$, the magnetic moment vanishes and $r^D(n) \to \infty$. 
Hence, the observation of the rings requires a fine tuning of the temperature ${\cal T}$ and of the dot-leads couplings $t_c$.
 
   In summary, a set of thermally induced interference rings can be seen when a quantum dot biased around the middle of a Kondo valley is studied above 
${\cal T}_K$ with a scanning gate. We have studied a case where the interaction effects are taken into account and which looks realistic enough 
for being amenable to experimental checks. We believe that the rings can be observed when the contact is biased between two resonances. This belief is 
supported by the numerical check that a contact made of two coupled sites in series (double dot setup) exhibits similar rings when it is biased between its 
two resonances. One motivation of this work comes from the interference ring observed~\cite{ensslin} using a QPC biased between two channel openings. We 
believe likely that this is due to similar interference effects, two sharp consecutive channel openings of a QPC playing~\cite{alp,lp} a similar role 
than the two consecutive resonance peaks considered in this work. Numerical studies are in progress for confirming this hypothesis. For the Kondo dot, 
the extension of the study below ${\cal T}_K$ and beyond the mean-field approximation is also in progress.  

\acknowledgments{This research has been supported by the EU Marie Curie network ``NanoCTM'' (project no.234970). 
Discussions with B. Brun, K. Ensslin, M. Sanquer and H. Sellier about SGM experiments are gratefully acknowledged.} 

\bibliography{KFP}
\end{document}